\documentclass[aps,prl,amsmath, twocolumn,showpacs,superscriptaddress,groupedaddress]{revtex4}  
\usepackage{graphicx}  
\usepackage{dcolumn}   

\usepackage[mathscr]{euscript}
\usepackage{amsmath}
\usepackage{amsfonts}
\usepackage{float}
\floatstyle{boxed}
\usepackage{color}
\usepackage{array}
\usepackage{cases}
\usepackage{comment}
\usepackage{subfigure}
\usepackage[usenames,dvipsnames,svgnames,table]{xcolor}

\begin{document}

\title{Shape from sound: toward new tools for quantum gravity}

\author{David Aasen} \affiliation{Department of Applied Physics, California Institute of Technology, Pasadena, CA , 91125}
\author{Tejal Bhamre} \affiliation{Department of Physics, Princeton University, Princeton, NJ 08544}
\author{Achim Kempf}  \affiliation{Departments of Applied Mathematics and Physics, University of Waterloo, Waterloo, N2L 3G1, Canada}\affiliation{Centre for Quantum Computing Technology, Department of Physics, University of Queensland, St. Lucia, Queensland 4072, Australia.}

%
%

%
\vskip 0.25cm
\date{\today}

\begin{abstract}
To unify general relativity and quantum theory is hard in part because they are formulated in two very different mathematical languages, differential geometry and functional analysis.  A natural candidate for bridging this language gap, at least in the case of the euclidean signature, is the discipline of spectral geometry. It aims at describing curved manifolds in terms of the spectra of their canonical differential operators. As an immediate benefit, this would offer a clean gauge-independent identification of the metric's degrees of freedom in terms of invariants that should be ready to quantize. However, spectral geometry is itself hard and has been plagued by ambiguities. Here, we regularize and break up spectral geometry into small finite-dimensional and therefore manageable steps. We constructively demonstrate that this strategy works at least in two dimensions. We can now calculate the shapes of 2-dimensional objects from their vibrational spectra.  
\end{abstract}

\maketitle

\pagenumbering{arabic}

The much sought-after unifying theory of quantum gravity, see e.g.,  \cite{QGtexts}, should also unify or at least very closely link the mathematical languages of quantum theory and gravity, namely functional analysis and differential geometry. A natural candidate for this mathematical link is the discipline of spectral geometry. Spectral geometry asks, for example, if or to what extent the shape of a thin object such as a vase can be described through its vibrational spectrum. For quantum gravity, the more general spectral geometric question is if or to what extent the curvature of compact $n$-dimensional Riemannian manifolds without boundaries (which henceforth we will simply call `manifolds') can be described in terms of the spectra of canonical differential operators on the manifold, such as the Laplace or Dirac operators. (Compactness here merely serves as an infrared cutoff that ensures that the spectra are discrete). Note that it is a different branch of spectral geometry that was popularized by M. Kac's paper \it Can one hear the shape of a drum? \rm \cite{Kac}. It asked to what extent the boundary of a flat membrane is determined by its spectrum, see e.g., \cite{Weyl,Pleijel,almost}.

If indeed at least euclidean spacetime curvature can be described in terms of the spectra of canonical operators, such as the Laplace or Dirac operators, this could be very useful for quantum gravity. This is because these spectra are invariants, i.e.,  they depend only on the Riemannian structure itself but not on gauge choices such as charts or frames. Therefore, the degrees of freedom of gravity would be cleanly identified as the relevant set of eigenvalues. Their dynamics and quantization would be free of difficulties stemming from diffeomorphism invariance, see e.g., \cite{Rovelli-Landi}. Even if the spectra uniquely determine the curvature only in a finite neighborhood of each manifold this could be very useful for the perturbative quantization of gravity. 

Concretely, spectral geometry studies maps, $S$, which map a set of curved manifolds into one or several spectra of canonical differential operators on the manifolds. Since any such $S$ is highly nonlinear, the study of the invertability of $S$ is hard. In order to make the calculations feasible, previous work has therefore generally  involved stringent assumptions, usually assumptions of high symmetry of the manifolds in question.  There have been intriguing results of spectra determining the curvature, as well as examples of isospectral non-isometric manifolds. For a review, see, e.g., \cite{haz}. Little is known, however, about the case that would be of most interest in physics, namely the generic case of manifolds without symmetries.

In \cite{Kempf1,AK-NJP}, it was proposed to make this nonlinear problem tractable by breaking it up into an iteration of linearized problems. To this end, the idea is to consider the derivative $DS$ of $S$. It maps small changes of the metric to the corresponding small changes of the spectra, a problem which is solvable by perturbation theory. If $DS$ is indeed invertible, one can then iterate small steps of curvature reconstruction to obtain for any given spectra the corresponding curved manifold(s). The existence of non-isometric but isospectral manifolds would then manifest itself in a path dependence of the manifold so obtained.

Regarding the question of the existence of isospectral non-isometric manifolds, we remark, that, crucially, $DS$ acts on changes to the metric that have generally to be assumed to come in scalar, vector and tensor types (as do, e.g., cosmological perturbations). This means that the spectra of the Laplacians on scalars, vectors and tensors will need to be considered so that $DS$ can possibly be invertible. This fact had been overlooked before the paper \cite{AK-NJP} and it explains why the Laplacian on scalars was often found to be insufficient to capture the metric.  

Here, we will now address the key question if $DS$ is indeed invertible in the generic case and if, therefore,
the spectra uniquely determine the curvature in a finite neighborhood of each manifold. We begin with the intuition that $DS$ can only be invertible if the number of degrees of freedom in perturbations of the metric at most matches the number of degrees of freedom in the spectra, except that both numbers are infinite, of course. To make the problem tractable, we propose to regularize the manifolds in question by approximating them through suitable geometric graphs embedded in a higher-dimensional manifold. This yields an $S$ that maps from a finite-dimensional space of geometric graphs into the finite-dimensional space of spectra of Laplacians defined over the graphs. The basic problem therefore reduces to determining when the finite-dimensional matrix $DS$ is invertible, while ensuring that the spectrum of the continuum Laplacian(s) is recovered in the continuum limit. 

While this program for constructing $DS^{-1}$ is relatively straightforward in principle, it is technically quite involved. This is because there are scalar, vector and tensor degrees of freedom in the perturbations of the metric. This means that our program generally involves Laplacians that act on the analogs of scalar, vector and tensor fields over geometric graphs that approximate manifolds. The theory of such Laplacians has not yet been fully developed, see e.g., \cite{desbrunetal}. 

However, we can immediately carry out our program on 2-dimensional manifolds. Their metric possesses only scalar degrees of freedom and therefore also the metric perturbations of 2-dimensional manifolds can be described by scalar functions. Also, all scalar functions can be expanded in the eigenbasis of the Laplacian on scalar functions. Therefore, crucially, there is exactly one shape degree of freedom for each eigenvector of the Laplacian. This is as many as there are eigenvalues, counting degeneracies. Notice that, for higher-dimensional manifolds, by the same argument, the number of shape degrees of freedom again cannot exceed the number of eigenvalues. We will now show that $DS$ is indeed generally invertible and also that the infinitesimal steps can be iterated. This then allows us to calculate shape from sound. 

\bf Construction of the geometric graphs. \rm  Our aim is to explicitly implement the above strategy for 2-dimensional manifolds. The first step is to regularize the problem to finite dimensions by approximating the manifolds by surfaces of polyhedra. To this end, assume that from the origin of $I\!\!R^3$ a large number, $N$, of half lines emanate in uniformly randomly chosen directions. Now let us assume further that $N$ positive numbers $\{r_i\}_{i=1}^N$ have been chosen. For each $i=1...N$ we mark on the $i$'th half line the point, $p_i$ which is at the radial distance $r_i$ from the origin. In this way, we obtain $N$ points $p_i$ in $I\!\!R^3$. Here, $p_i = r_i e_i$ and $e_i$ are normalized direction vectors. Further assume a triangulation of the set of points $\{p_i\}$ so that we obtain the surface of a polyhedron. 
Any star-shaped Riemannian manifold can be approximated to arbitrary precision by choosing $N$ sufficiently large. Generalizations of this construction for non star-shaped manifolds should be possible similarly. 

We will work with arbitrarily high but finite precision. To this end, let us fix a large $N$, a set of random directions of the $N$ half lines and a triangulation. We then allow only the $N$ values $\{r_i\}$ to vary. In this way, we obtain an $N$-parameter family of polyhedra that approximate star-shaped manifolds. The key question then becomes whether from the vibrational spectrum of these polyhedral surfaces one can tell their shape. 

In order to answer this question, we first note that the number of real degrees of freedom that describe the shape of a polyhedron in the family is the number of radii, $r_i$, of which there are $N$. We need to compare this number with the number, $N'$, of eigenvalues $\{\lambda_j\}_{j=1}^{N'}$ of the Laplacian of the surface of the polyhedron. While there are several definitions for Laplace operators on graphs, they all possess the same number, $N'$, of eigenvalues, namely $N'=N$. 

To see this, note that functions on the manifold are replaced on the graph by functions on the $N$ nodes $\{p_i\}_{i=1}^N$, see, e.g., \cite{Fan}. The function space ${\cal{F}}$ is  therefore an $N$-dimensional Hilbert space. The Laplace operator, $L$, on functions over the graph acts on ${\cal{F}}$ which means that it is a self-adjoint $N \times N$ matrix . The spectral theorem then implies that $L$ possesses $N'=N$ real eigenvalues.

{\bf The geometric graph Laplacian.} 
In order to study the information contained in the spectrum in detail, we need to choose an explicit graph Laplacian. The choice is not unique, much like there is no unique discretization of a first derivative. General requirements are that it must be self-adjoint and semi-positive definite \cite{Rosenberg}. Here, for the triangular surface meshes that we work with, we also need locality, linear precision and positivity of the weights \cite{Wardetzky}. Choices of discrete geometric Laplacians are discussed in \cite{Reuter}. We will use the standard discrete geometric Laplacian for triangulations introduced by Desbrun et al. \cite{Desbrun}. It obeys these conditions and is also widely used in computer graphics. 

To define it, recall that a graph, $G$, of a polyhedron consists of a set $V$ of nodes or vertices, $v_i$, given by the points $p_i\in I\!\!R^3$, and a set of weighted edges $E$. By $d_v$ we denote the degree of the vertex $v$, i.e., the number of edges emanating from it. The weight matrix $W$ has as matrix elements the weights $w_{ij}$ that are associated with the edge that joins the vertices $i$ and $j$:
\begin{equation}
w_{ij}=\frac{\cot(\gamma_{ij})+\cot(\beta_{ij})}{2},
\end{equation}
Here, $\gamma_{ij}$ and $\beta_{ij}$ denote the two angles opposite the edge $(i,j)$. 
The Laplacian is then defined through $L=V-W$, where $V=diag(v_1, v_2, \ldots, v_n)$, with $v_i=\sum_{j}w_{ij}$.  

Crucially, this graph Laplacian is also known to be geometric in the sense that it converges to the ordinary Laplacian in the continuum limit. Concretely, the results of \cite{convergence} show that if a set of so-called normal polyhedra converges in distance and in area then the geometric graph Laplacian converges to the Laplacian on the manifold with respect to the operator norm topology. The theorem applies because, for sufficiently large $N$, our star-shaped polyhedra will be normal graphs in the sense of \cite{convergence}, meaning that every point on the polyhedra has a unique closest point on $M$, whose distance is smaller than a fixed upper bound (the so-called reach of $M$). This means that given a star-shaped compact Riemannian manifold $M$, our polyhedral approximations possess spectra that converge to the spectrum of the Laplacian on the manifold for $N\rightarrow\infty$. 

Now as is easy to verify,  zero is always an eigenvalue of this graph Laplacian, with the constant vector being the eigenfunction (as on the manifolds). Thus, only the eigenvalues $\{\lambda_i\}_{i=2}^N$ can carry shape information (although if we allowed nontrivial topologies of the polyhedra, the degeneracy of the eigenvalue $0$, being the $0$'th Betti number, would carry connectivity information). Also, the graph Laplacian is conformally invariant since  it depends only on the angles between edges but not on their lengths, like the conformal Laplacian on the manifold.  Therefore, without restricting generality, only the radii $\{r_i\}_{i=2}^N$ carry shape information, while $r_1$ fixes the overall size. The numbers of degrees of freedom of shape and sound appear therefore matched at $N-1$. 

{\bf Infinitesimal Inverse Spectral Geometry of Graphs.}
In order to determine if, therefore, small changes of sound indeed determine the corresponding small changes of shape, let us consider the nonlinear map $S :I\!\!R^{N-1} \rightarrow I\!\!R^{N-1}, $ namely $S: (r_2,...,r_N) \rightarrow (\lambda_2,...,\lambda_N)$. Its derivative, $DS$, with entries $\partial \lambda_i/\partial r_j$ is a $(N-1)\times(N-1)$ matrix.  $\det(DS)$ is a generic function and should therefore be nonzero almost everywhere (in the sense of on a set of measure zero), so that  $(DS)^{-1}$ should exist. Almost all small changes of a polyhedron's spectrum should uniquely determine a corresponding change of shape. 

Concretely, since the graph Laplacian is conformally invariant, we fix one point on the graph, $r_1$. As discussed, this fixes the overall size of the polyhedron and leaves us with only $N-1$ radii $\{r_i\}_{i=2}^N$ to determine from the spectrum, which has $N-1$ variable eigenvalues $\{\lambda_i\}_{i=2}^N$. Now suppose the shape is changed by slightly changing the radii $\{r_i\}_{i=2}^N$:
\begin{equation}
r_i \rightarrow r_i +\epsilon a_i, ~~~i=2,...,N
\end{equation}
where $\epsilon \ll 1$ and all $a_i \in I\!\!R$. Then, 
\begin{equation}
L \rightarrow L + \epsilon \Lambda(a), ~~~\mbox{with}~~~a = (a_2,...,a_N)
\end{equation}
We find that $\Lambda$ = $\delta V$ - $\delta W$, where $\delta v_i=\sum_j \delta w_{ij}$, with:
 \begin{equation}
 \delta w_{ij} =\eta_{\hat{\gamma}ij}\cot{\gamma_{ij}}+\eta_{\hat{\beta}ij}\cot{\beta_{ij}}
 \end{equation}
The indices $\hat{\gamma}$ and $\hat{\beta}$ denote the number of that vertex opposite the edge $(i,j)$ which has the angle $\gamma_{ij}$ or $\beta_{ij}$ respectively. Here,
$
\eta_{kij} = (l_{ik} \cdot  f_{ik} +l_{jk} \cdot f_{jk})/(l_{ik} \cdot  l_{jk}) 
$
with $l_{ij} = r_i e_i-r_j e_j$ and $f_{ij} = a_i e_i - a_j e_j$. The change in the spectrum of $L$ can now be calculated by first order perturbation theory.
 With $k^{(i)}$ denoting the $i$'th eigenvector of $L$
\begin{equation}
\lambda_i \rightarrow \lambda_i + \epsilon \sum_{j_1,j_2=1}^N k^{(i)}_{j_1}\Lambda_{j_1,j_2}(a)k^{(i)}_{j_2},
\end{equation}
i.e., the linear map $DS$ reads
\begin{equation}
(DS)_{i,j} =\partial \lambda_i/\partial r_j = \langle k^{(i)}\vert \Lambda(a^{(j)})\vert k^{(i)}\rangle
\end{equation}
where $(a^{(j)})_k=\delta_{j,k}$ is the perturbation vector in the $j$-direction. 
For any shape, i.e., for any point $r=(r_2,...,r_N)$ in the $(N-1)$-dimensional space of radii, ${\mathscr{R}}$, this yields $DS(r)$. Its determinant, $\det(DS(r))$ is a real-valued function over $\mathscr{R}$.
If there is no further loss of degrees of freedom (such as that the zero eigenvalue never moves), the single equation $\det(DS(r))=0$ should describe a submanifold $\mathscr{Z} \subset \mathscr{R}$ which is at most $(N-2)$-dimensional. This then means that $DS(r)$ is invertible almost everywhere, namely on $\mathscr{R} - \mathscr{Z}$. Notice that spectra may uniquely determine shapes even on $\mathscr{Z}$, similarly to how $g: x \rightarrow x^3$ is invertible for all $x$, including $x=0$, in spite of $g'(0)=0$.

To test that there is no further loss of degrees of freedom (other than the conformal invariance and the constancy of the zero eigenvalue that we took care of) we should evaluate the determinant for all $r\in \mathscr{R}$. The explicit form of $\det(DS(r))$ is unwieldy due to its dependence on eigenvectors. It is straightforward, however, to evaluate $\det(DS(r))$ numerically for any $r\in\mathscr{R}$. The results indicate that the determinant is indeed nonzero almost everywhere. This implies, by the inverse function theorem, that at least for almost all the polyhedra there is a finite neighborhood within which any change of shape is recoverable from its change of spectrum. 

{\bf From infinitesimal to finite inverse spectral geometry.}
The fact that $DS(r)$ is invertible almost everywhere in the space of shapes, $\mathscr{R}$, suggests to tackle finite inverse spectral geometry by iterating the infinitesimal steps. For example, a simple strategy for finding a target polyhedron when given only its spectrum is to start with an arbitrary ansatz polyhedron. Then, one slightly perturbs its spectrum towards the target spectrum, while calculating the corresponding change of shape. The procedure is iterated on a straight line path in the space of spectra, all the while calculating the change of shape until the target spectrum is reached. This works as long as $\det(DS(r))\neq 0$ on this path, i.e., until $\mathscr{Z}$ needs to be crossed. In this case, one strategy is to temporarily reduce the number of radii and eigenvalues that one varies. The resulting lower-dimensional submatrix of $DS(r)$ can be chosen to have nonzero determinant and is therefore invertible. Once $\mathscr{Z}$ is crossed, one can return to the straight path to the target in the space of spectra.  

Another strategy can be devised that is fundamentally more robust when $\mathscr{Z}$ needs to be crossed: 
We start again with an arbitrary ansatz graph, given by a set of radii $\{r_i\}_{i=2}^N$. For this algorithm, we use a distance function, $\sigma$, that measures the distance between this graph's spectrum and the spectrum of the target graph. We perturb the graph, i.e., its radii, in the direction of the gradient $\frac{\partial \sigma}{\partial r_i}$. By iteration we obtain in the space of shapes, $\mathscr{R}$, the steepest path to the target spectrum and shape, with respect to the chosen notion of distance. 

This algorithm does not automatically halt when crossing $\mathscr{Z}$. To see this, for any graph, specified by radii $r=\{r_i\}$, we define its spectral distance $\sigma$ to the target graph through
\begin{equation}
\sigma(r) := \sum_{i=2}^N \left(f(\lambda_i(r))-f(\lambda_i^{target})\right)^2,
\end{equation}
where $f$ is some arbitrary but fixed strictly monotonic function. More general $\sigma$ are possible, of course, but this level of generality suffices. 

The only possibility for this algorithm to halt is when the gradient vanishes, i.e., when $\partial \sigma/\partial r_i=0$ for $i=2,...,N$. This happens when the global minimum of $\sigma(r)$ is reached at the target $r=r^{target}$ or at an isospectral graph if one exists. It can also happen at other critical points of $\sigma(r)$, where for all $j=2,...,N$:
\begin{eqnarray}
 0 = \frac{\partial \sigma}{\partial r_j} &=& \sum_{i=2}^N 2\left(f(\lambda_i({r}))-f(\lambda_i^{target})\right)\frac{\partial f(\lambda_i({r}))}{\partial \lambda_i}\frac{\partial \lambda_i({r})}{\partial r_j}  \hspace{2mm} \nonumber \\
   &=& \sum_{i=2}^N v_i \frac{\partial \lambda_i({r})}{\partial r_j} = 
\sum_{i=2}^N v_i (DS)_{i,j}  
\end{eqnarray}
Here, $v_i =  2(f(\lambda_i({r}))-f(\lambda_i{}^{{target}}) \frac{\partial f(\lambda_i)}{\partial \lambda_i}$. Therefore, the algorithm can halt prematurely at a $r\in \mathscr{Z}$, but only if $v(r)$ also happens to be  in the kernel of $(DS)^T$. In this case, the algorithm can be unstuck by changing the definition of $f$, because this changes  $v(r)$. Intuitively, the algorithm flows down the landscape $\sigma(r)$ over $\mathscr{R}$ along the gradient. It halts if it runs into a local minimum of $\sigma(r)$. By changing $f$ one can change the landscape so that the minimum and its catchment area are removed from $r$. Alternatively, one can keep $f$  fixed and randomly move $r$ sufficiently to escape the minimum's attractor.  

{\bf Numerical results.}
We implemented the gradient-based algorithm, choosing the half lines' directions randomly or regularly. We choose $f: x\rightarrow 1/x$, so that the algorithm tends to first converge the smaller eigenvalues, which determine the large scale shape, and then fixes the small scale structure.  Technically, for large $N$, there exist polyhedra that are near translates and rotates of the target polyhedron and that therefore would be near isospectral, slowing down convergence. This is easily fixed by fixing several radii instead of just one. 

We have found that the algorithm allows one to recover graphs from their spectra all the way down to machine precision. In fact, in the generic case, i.e., when choosing $N$ random directions for the half lines, we have not found any iso-spectral non-identical graphs.  Else, when the half-lines are chosen symmetrically, we do of course have that at least each graph and its mirror image possess the same spectrum. 

\begin{figure}[htbp]
	\begin{center}
	 \includegraphics[width=3.2in]{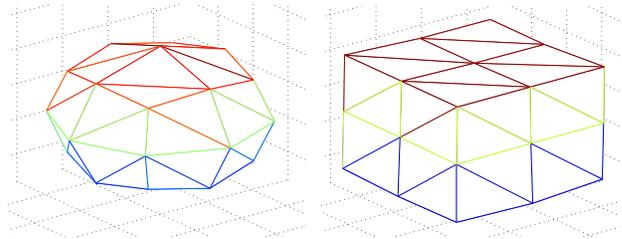}
	\end{center}
	\caption{Snapshots of the algorithm starting with a sphere and finding the cube from its spectrum alone.}
	\label{fig:blah-vs-blah}
\end{figure}

{\bf Conclusions and Outlook.} Methods that enable one to fully describe manifolds in terms of the spectra of their canonic operators should have great potential for quantum gravity. This is because these eigenvalues, being diffeomorphism-independent invariants of the manifold, can then be identified as the pure, gauge-independent degrees of freedom of gravity. If applied to spacelike hypersurfaces, their dynamics may be quantized along the lines outlined for the Dirac operator in \cite{Rovelli-Landi}. Applied to 4-dimensional manifolds, euclidean quantum gravity can then be pursued by studying the thermodynamics of these spectra. In fact, the euclidean Einstein action in four dimensions is known to be expressible very simply in terms of the eigenvalues of the Laplacian, \cite{AK-NJP,Hawking}. 

In \cite{AK-NJP}, it was proposed, therefore, to break up the nonlinear problem of spectral geometry into the iteration of infinitesimal linear steps. Here, we have made this idea workable, at least for two-dimensional star-shaped manifolds, by introducing a program of graph regularization. This showed analytically that small changes of spectrum generically do determine the small changes of shape. The numerics even indicates that, in the absence of symmetry, the spectrum determines the shape globally.   

In order to extend our results to four (euclidean) dimensions, it will be necessary to further develop the theory of Laplacians on tensors on geometric graph approximations of higher-dimensional manifolds. But it may also be possible to pursue a similar program with ultraviolet regularizations other than graph regularizations, so as to preserve the smooth manifold structure throughout.

\bf Acknowledgements: \rm AK gratefully acknowledges the kind hospitality at the University of Queensland during his sabbatical visit, as well as support by the Canada Research Chairs and Discovery programs of NSERC. DA gratefully acknowledges the support of the NSERC-PGSM program. The authors gratefully acknowledge useful discussions in particular with William Donnelly, Mikhail Panine, Ingo Roth and Laurel Stephenson-Haskins.


\begin{thebibliography}{10}

\bibitem{QGtexts} G. Gibbons, S. W. Hawking, \it Euclidean Quantum Gravity, \rm 
World Scientific Publishing Co. Ptd. Ltd., Singapore (1993), 
  C. Rovelli, \it Quantum Gravity, \rm Cambridge Monographs on Mathematical Physics, Cambridge University Press, Cambridge (2004), C. Kiefer, \it Quantum Gravity, \rm Oxford University Press, Oxford (2007)
\bibitem{Kac}M. Kac, Am. Math. Monthly, \textbf{73}, 1 (1966).
\bibitem{Weyl}H. Weyl, Nachr. d. K\"onigl. Ges. d. Wiss. zu G\"ottingen, {\bf 1,} 110 (1911).
\bibitem{Pleijel}  S. Minakshisundaram and \r{A}. Pleijel, Can. J. Math. \textbf{1}, 242 (1949).
\bibitem{almost}   M. Ben Haj Rhouma, M. A. Khabou, L. Hermi, 
\it Adv. Imaging \& Electron Physics, \rm 
{\bf 167}, 185 (2011).
\bibitem{Rovelli-Landi}G. Landi, C. Rovelli, Phys. Rev. Lett. \textbf{78}, 3051 (1997).
\bibitem{haz}K. Datchev and H. Hezari, \it Inverse Problems and Applications: Inside Out II, \rm Math. Sci. Res. Inst. Publ. Series, \textbf{60}, Section 10, Cambridge University Press, Cambridge (2012).
\bibitem{Kempf1} A. Kempf, Phys. Rev. Lett. \textbf{85}, 2873 (2000).
\bibitem{AK-NJP} A. Kempf, New J. Phys. \textbf{12}, 115001 (2010).
\bibitem{desbrunetal} M. Desbrun, A. N. Hirani, M. Leok, J. E. Marsden, arxiv:math/0508341  
\bibitem{Fan} F. Chung. \emph{Spectral Graph Theory}, Vol. 92 of CBMS Regional Conference Series in Math., American Mathematical Society, Washington (1997).
\bibitem{Rosenberg}S. Rosenberg, \emph{The Laplacian on a Riemannian manifold} Cambridge University Press, Cambridge (1997).
\bibitem{Wardetzky} M. Wardetzky, S. Mathur, F. K\"alberer and E. Grinspun, Eurographics Symposium on Geometry Processing, Eds. B. Alexander, G. Michael, 
ACM International Conference Proceeding Series \textbf{257}, 33 (2007).
\bibitem{Reuter}M. Reuter et al. Computers \& Graphics,  \textbf{33} (3), 381 (2009).
\bibitem{Desbrun} M. Desbrun et al., in \it Proc. 26th annual conf. on computer graphics and interactive techniques, \rm Proceedings of the ACM Siggraph, Addison-Wesley Publishing Co., New York, NY \rm p. 317 (1999).
\bibitem{convergence} K. Hildebrandt, K. Polthier, M. Wardetzky, Geometriae Dedicata, {\bf 123} (1), 89 (2006). 
\bibitem{Hawking}S. W. Hawking, Phys. Rev. D \textbf{18}, 1747 (1978).






\end{thebibliography}
\end{document}